\documentclass[12pt,conference]{IEEEtran}
\usepackage{fancyhdr}
\usepackage{leading}
\leading{12pt}

\usepackage{amsmath, amssymb, cite, epsfig}
\usepackage{graphicx}
\usepackage{dblfloatfix}
\usepackage{array}
\usepackage{textcomp}
\newcolumntype{P}[1]{>{\centering\hspace{0pt}}p{#1}}
\newcolumntype{M}[1]{>{\centering\hspace{0pt}}m{#1}}
\newcolumntype{L}{>{\centering\arraybackslash}m{3cm}}

\usepackage{multirow}
\usepackage[usenames,dvipsnames,table]{xcolor}

\usepackage{filecontents}
\usepackage{enumerate}
\graphicspath{ {figures/} }

\graphicspath{{figures/}}

\usepackage{etoolbox}
\makeatletter
\patchcmd{\@makecaption}
{\scshape}
{}
{}
{}
\makeatletter
\patchcmd{\@makecaption}
{\\}
{.\ }
{}
{}
\fancypagestyle{firststyle}{
	\fancyhf{}
	\fancyhead[L]{ O. Kanhere and T. S. Rappaport, ``Position Location for Futuristic Cellular Communications - 5G and Beyond,'' \textit{in IEEE Communications Magazine}, vol. 59, no. 1, pp. 70-75, January 2021.}     %

}
\makeatother
\definecolor{TableColor}{RGB}{217,217,217}
\begin{document}
	\rowcolors{1}{TableColor}{white}
\title{ Position Location for Futuristic Cellular Communications - 5G and Beyond}
\author{\IEEEauthorblockN{Ojas Kanhere and Theodore S. Rappaport\\}
	
\IEEEauthorblockA{	\small NYU WIRELESS\\
					NYU Tandon School of Engineering\\
					Brooklyn, NY 11201\\
					\{ojask, tsr\}@nyu.edu}}\vspace{-0.7cm}

\maketitle
\thispagestyle{firststyle}

\begin{abstract}
With vast mmWave spectrum and narrow beam antenna technology, precise position location is now possible in 5G and future mobile communication systems. In this article, we describe how centimeter-level localization accuracy can be achieved, particularly through the use of map-based techniques. We show how data fusion of parallel information streams, machine learning, and cooperative localization techniques further improve positioning accuracy.

\end{abstract}

\section{Introduction}\label{Introduction}
Precise position location (also called \textit{positioning} or \textit{localization}) is a key application for the fifth generation (5G) of mobile communications and beyond, wherein the position of objects is determined to within centimeters. With the rapid adoption of Internet of Things (IoT) devices, a variety of new applications that require centimeter-level precise positioning shall emerge, such as automated factories that require precise knowledge of machinery and product locations to within centimeters. Geofencing is the creation of a virtual geographic boundary surrounding a region of interest to monitor people, objects, or vehicles, and by using sensors on a moving object, the location of the object may be continually and adaptively ``geofenced" to trigger a software notification immediately when the object enters or leaves the virtual geographic boundary. Position location to within 1-2 m will enable accurate geofencing, such that users entering/leaving a room or equipment and people may be tracked in hospitals, factories, within and outside buildings.

Today's fourth generation (4G) cellular networks rely on  LTE signaling and the global positioning system (GPS) (which is accurate to within 5 m). However, in indoor obstructed environments, or in underground parking areas and urban canyons, GPS signals are attenuated and reflected such that user equipment (UE) cannot be accurately localized. To further refine the positioning capabilities of GPS indoors and in urban canyons, SnapTrack ``\textit{wireless assisted GPS}" (WAG) improved the sensitivity of GPS receivers. Additionally, databases of geo-tagged Wi-Fi hotspots have been used by companies such as Apple and Google. The UE may be localized using the known positions of all Wi-Fi hotspots that the UE can hear, where the UE position estimate is formed from the weighted average of the received signal strengths, providing an accuracy of tens of meters. Although FCC requirements specify a horizontal localization error of less than 50 m for 80 percent of enhanced 911 (E911) callers, a localization error less than 3 m will be required for positioning applications of the future. Additionally, FCC requires a vertical localization error less than 3 m for 80 percent of E911 callers by April 2021, to identify the caller's floor level, which is achievable using barometric pressure sensors present in modern cell phones (see FCC's Fifth Report and Order PS Docket 07-114.).

In addition to infrastructure-based positioning systems, other sensor-based technologies such as vision-based localization using cameras (commonly utilized by drones\cite{Pedersen_2018}) can provide accurate positioning capabilities when fused with inertial sensors. However, in low-visibility environments, localization systems at cellular frequencies work better since they are not blocked when visibility is hampered. Ultrasound indoor positioning systems such as Forkbeard are able to achieve a precision level of 10 cm within an office environment. Autonomous vehicles utilize light detection and ranging (LIDAR) to estimate the relative distances to other vehicles with sub-millimeter accuracy \cite{guan2016}, while factory-based systems using infrared have shown good accuracy \cite{Rap89b}. 

Position location solutions are being developed using other media such as ultra wideband (UWB), RFID, visible light, and Bluetooth. UWB signals, in the 3.1-10.6 GHz band, have a bandwidth of more than 500 MHz. Rapid strides in utilizing UWB for localization are expected, with the iPhone 11 currently carrying UWB chips that are typically capable of achieving a ranging accuracy on the order of centimeters \cite{Falsi_2006}.

The advent of millimeter-wave (mmWave) communications enables a paradigm shift in localization capabilities by allowing joint communication and position location, utilizing the same infrastructure. As shown in this article, the massive bandwidths, coupled with the high gain directional, steerable multiple-input multiple-output (MIMO) antennas at mmWave frequencies, enable unprecedented localization accuracy in smartphones of the future. We demonstrate how the utilization of cooperative localization, machine learning, user tracking, and multipath enables precise centimeter-level position location.

\section{Fundamental Localization Techniques}\label{Present}

Today's localization solutions primarily focus on geometric localization with augmented assistance, wherein the position of the base station (BS) is known and the UE location is determined based on geometric constraints such as the BS-UE distances and physical angular orientations between BS and UE. 

\begin{figure}
	\centering
	\includegraphics[width=0.45\textwidth]{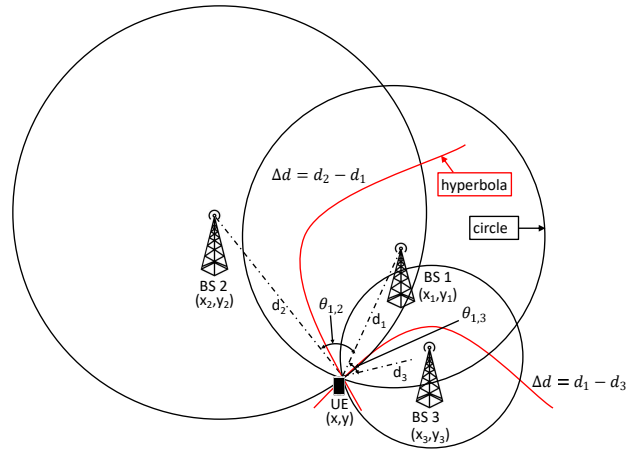}
	\caption{ The UE may be localized based on ToA (black circles), TDoA (red hyperbola), or AoA (black dotted lines) localization techniques \cite{Rap96a}.}
	\label{fig:all_methods}
\end{figure}

In angle of arrival (AoA) localization technique, the UE estimates the angle of the strongest received signal. AoA positioning was conceived for E911 in the early days of cellular\cite{Rap96a}. In time of arrival (ToA) (or time difference of arrival, TDoA) localization techniques, the UE estimates the distance (or difference in distance) from the BS by estimating travel time (or differences in travel time) of the reference signal from the BS. The UE may then be localized to the point where the circles (or hyperbolas) corresponding to the BS-UE distances intersect.  A spatial resolution of up to 2.44 m and 4.88 m is achievable with 5G New Radio (NR) waveforms for ToA and TDoA measurements, respectively \cite{3GPP.36.133}. In addition to utilizing GPS for UE localization, 4G (and future 5G) networks implement TDoA localization and utilize the barometric pressure sensors in UE for altitude estimation\cite{Pedersen_2018}. The operation of AoA, ToA, and TDoA localization techniques is illustrated in Fig. \ref{fig:all_methods} and is well understood. 

\subsection{Accurate Localization in 5G Networks with Directional Antenna Arrays and Wide Bandwidths}
In the 5G era, it is now possible to achieve very accurate localization performance with highly directional antenna arrays having narrow beamwidths and wide bandwidths \cite{Rappaport19a}. The frequency range (FR) 2 of 5G NR covers mmWave frequencies ranging from 24.25 GHz to 52.6 GHz. Additionally, the IEEE 802.11 ad standard supports the use of the 60 GHz mmWave band indoors, from 57 GHz to 71 GHz.

\begin{figure}
	\centering
	\includegraphics[width=0.5\textwidth]{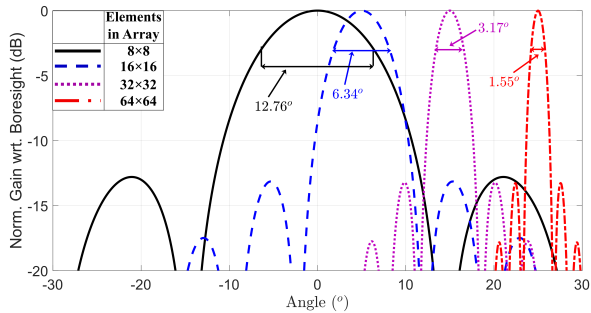}
	\caption{ The normalized antenna gain (with respect to boresight - the axis of maximum gain) of URAs with 8$ \times $8, 16$ \times $16, 32$ \times $32, and 64$ \times $64 array elements. Note the half power beamwidths (HPBWs) are 12.76\textdegree,  6.34\textdegree, 3.17\textdegree,  and 1.55\textdegree\ respectively.  }
	\label{fig:angular_accuracy}
\end{figure}

The short wavelength in the mmWave frequency band allows electrically large (but physically small) antenna arrays to be deployed at both the UE and BS. MmWave BS antenna arrays with 256 antenna elements and 32-element mobile antenna arrays are already commercially available. The frequency-independent half-power beamwidth (HPBW) of a uniform rectangular array (URA) antenna with half-wavelength element spacing is approximately (102/N)\textdegree{}, where $ N $ is the number of antenna elements in each linear dimension of the planar array \cite{Balanis_2016}, as seen in Fig. \ref{fig:angular_accuracy}. 

Narrower HPBWs of antenna arrays allow the AoA of received signals to be estimated precisely, and further signal processing provides better accuracy. For example, the sum-and-difference for an infrared system technique achieved sub-degree angular resolution with two overlapping and slightly offset antenna arrays \cite{Rap89b}, showing it is possible to very accurately detect precise AoA at UEs or BSs.

Although mmWave frequencies suffer from higher path loss in the first meter of propagation and experience greater blockage losses compared to lower frequencies, the greater gain provided by the directional antennas coupled with smaller serving cells (100-200 m radius) compensates for the additional path loss. Indeed, recent research \cite{Gante_2020} demonstrates the feasibility of using mmWave for outdoor localization.

Utilization of mmWave frequency bands will enable unprecedented positioning accuracy due to the ultra-wide bandwidths available, since the larger bandwidths allow finer time resolution of multipath signals transmitted from the BS to the UE, on the order of a nanosecond, where a 1 ns time resolution implies a spatial resolution of 30 cm before additional processing that can further improve accuracy. 

\subsection{Performance of Fundamental Localization Techniques in Dense Multipath Environments}
ToA, TDoA, and AoA localization techniques were designed for line-of-sight (LoS) propagation. In indoor/outdoor non-line-of-sight (NLoS) environments however, multipath arrives at different angles with larger delays, yielding positioning error. Without using any advanced correction techniques, a poor mean error of 10 m was observed with well-known AoA localization based on NLoS indoor office measurements\cite{Kanhere18a}. Similar enormous mean errors of 8-10 m inside buildings were observed in NLoS when the localization performance was tested using traditional methods from outdoor E-911 \cite{Rap96a} via simulations in NYURay, a 3D mmWave ray tracer \cite{Rappaport19a}. The poor localization accuracy of known approaches, in the face of multipath and an obstructed or weak LoS signal, motivates the need to develop more accurate and robust localization approaches that exploit the wide bandwidth and narrow beamwidths of 5G and beyond for multipath-rich NLoS environments.

\subsection{NLoS Mitigation for Accurate Positioning}
To combat the poor performance of traditional ToA-, TDoA-, and AOA-based localization techniques in NLoS environments, NLoS mitigation techniques can identify and then discard NLoS signals to only use the LoS BSs for localization. This subsection describes a variety of techniques to selectively identify and discard the NLoS signals. 

In \cite{xiong13} the authors observed that with conventional WiFi radios operating at 2.4 GHz, the AoA was stable over small UE movements (5 cm) in LoS environments, while in NLoS environments, the AoA varied by more than 5\textdegree\ if the UE was moved by 5 cm. If the AoA of the received power varied by more than 5\textdegree\ when the UE was moved by 5 cm, the signal was assumed to correspond to an NLoS path and thus discarded from use in estimating position. By suppressing NLoS multipath and only using the LoS path, a median localization accuracy of 23 cm was achieved with six 2.4 GHz WiFi access points \cite{xiong13}.

Estimating the BS-UE distance, a critical step for ToA localization, may additionally be utilized to determine whether the BS-UE link is in NLoS. The running variance of the BS-UE distance estimates ($ \sigma^2 $) in NLoS is greater than LoS; hence, NLoS BS-UE links may be identified based on the running variance observed in real time. The UE can accurately be assumed to be in NLoS (and the UE-BS link is not used for localization) when $ \sigma^2 $ is greater than a calibrated threshold $ \gamma $ \cite{Schroeder_2007}. The variance of distance estimates is greater for a mobile user than for a stationary user due to the change in the true BS-UE distance when the UE is in motion. To account for user motion, $ \gamma $ must be increased, and in \cite{Schroeder_2007}, a constant proportional to the square of the velocity of the user was added to $ \gamma $ to account for user motion. 

Channel features such as maximum received power, root mean square (RMS) delay spread, Rician-K factor, and the angular spread of departure/arrival may be utilized to determine whether the UE is in NLoS \cite{Huang_2019}. NLoS channels typically have lower maximum received power over the power delay profile (PDP) due to the presence of obstructions and reflectors. The delay spread of multipath components is higher in NLoS environments. The K-factor of a channel is equal to the ratio of the square of the peak amplitude of the dominant signal and the variance in the channel amplitude and is known to indicate the degree of multipath in a signal \cite{Huang_2019}. In NLoS channels, due to the absence of a direct path, the K-factor is close to 0 dB. The angular spread of NLoS channels is wider since the multipath components arrive from varied directions.

NLoS classification accuracy is improved when multiple channel characteristics are used in tandem \cite{Huang_2019}. A support vector machine (SVM) is a popular classifier capable of classifying data based on multiple parameters. An SVM utilizes channel characteristics to determine a hyperplane, which divides data into two classes. For NLoS identification, the SVM determines the optimal hyperplane to divide data into LoS and NLoS classes. In \cite{Huang_2019}, an SVM was shown to outperform individual channel features, reducing the NLoS identification error rate from 10 percent to 5 percent.

\section{Sub-meter Precise Position Location}
Identifying and discarding NLoS signals to only use LoS signals for localization wastes multipath signal energy, and requires dense BS deployment since the UE must be in LoS of two or more BSs for classical LoS positioning techniques to work. However, such over-deployment of BSs may be cost-prohibitive. We shall now look at alternative localization techniques wherein the UE utilizes information from neighboring UEs, and exploits NLoS BSs, and multipath.

\subsection{Cooperative Localization}
With the introduction of device-to-device (D2D) communication protocols in 5G\cite{Pedersen_2018}, an exciting avenue for cooperative localization has opened up. UEs may now directly communicate with one another instead of communicating with the BS alone in order to achieve localization of all UE. 

Due to dedicated communication resources allocated for D2D communication in 5G, UEs may conduct range and angular measurements on each D2D link. Since UEs are typically located closer to one another than to BSs, the probability of D2D links being LoS and having higher signal-to-noise ratio (SNR) is greater, providing better positioning accuracy. In a network with $ N $ UEs, up to $ N\choose 2 $ additional D2D link measurements are possible. 

The relative UE location information, extracted from the D2D link measurements, may be sent to a central localization unit co-located at one of the serving BS or a central server (i.e., centralized cooperative localization). The position of all the UEs in the network is simultaneously determined by nonlinear least squares (LS) estimation, wherein the positions of the UEs that jointly minimize the deviation from the physical angular orientation and distance-based link constraints are determined. Optimization techniques such as the Levenberg-Marquardt algorithm (LMA) \cite{Shahmansoori2018}, which combines the Gauss-Newton algorithm and the method of gradient descent, may be used for nonlinear LS estimation.

Centralized cooperative localization in future dense IoT networks may lead to network congestion if all localization messages are routed to a central server. In distributed algorithms, UEs are localized based on local measurements exchanged by neighboring nodes (as is done in centralized localization). The location estimates of the UEs are then iteratively refined until all neighboring UEs reach an agreement \cite{Pedersen_2018}. While not as accurate as infrared methods [3], a root mean square error of 2.5 m and 3 m was achieved in an indoor environment with centralized and distributed cooperative localization, respectively, over an area of approximately 40 m $ \times $ 20 m with four BS with known locations and 13 unknown UE locations \cite{Pedersen_2018}.

\subsection{Machine Learning for Localization}
In contrast to geometry-based localization algorithms, machine learning provides a data-centric view of the UE localization problem. Localization algorithms that employ machine learning first create a ``fingerprinting database" of the environment during the training (offline) phase\cite{Gante_2020}. A fingerprint is a vector containing channel parameters such as the received signal strength (RSS), channel state information (CSI), and the AoA of the strongest signal of all BS links measured a priori at known locations called \textit{reference points}, distributed throughout the environment. A fingerprinting database is constructed by storing the fingerprint measured at each reference point with the coordinates of the reference point.

Once the fingerprinting database is constructed, then in the real-time online position location step the BS-UE channel is measured by the UE. The channel measurements are \textit{matched} to the fingerprinting database (stored in the UE or in the network) to determine the UE position. Matching may be done via maximum a posteriori (MAP) estimation.

Alternatively, matching may be performed by utilizing a similarity criterion to compare the online measurements to the fingerprinting database. A common similarity criterion is the distance, such as the Euclidean ($ L_2 $) or the Manhattan ($L_1$) distance, of the online measurements from the channel measurement at the reference points. In the k-nearest neighbor (k-NN) algorithm, the user position is the weighted average of the \textit{k} ``\textit{nearest}" reference points.

The UE localization problem can be restated as determining the nonlinear function that transforms the channel parameters into a position estimate. A neural network determines the nonlinear function, based on data available in the fingerprinting database. A neural network is a series of multi-level nonlinear functional transformations of the input, which can be used to approximate a target function. For user localization, the inputs to the neural network are the measured channel parameters, and the target function is the positional coordinates of the user. Successive layers of a neural network are combined linearly by weights. The optimal weights that transform the inputs (channel parameters) as close as possible to the target function (user position) are found in the offline training phase by minimizing the closeness of the output of the neural network to the target function at the reference points.

Machine-learning-based localization algorithms require the availability of a dense fingerprinting database, the creation of which is a time-intensive process. The localization accuracy of fingerprinting algorithms depends on the distance between reference points, with the localization accuracy typically on the order of the distance between the reference points. Additionally, changes in the environment such as the addition of new furniture require the fingerprinting database to be re-created. Transfer learning may be leveraged to reduce the amount of data required. Theoretical radio wave propagation models are leveraged to replace data collection partially by ray tracing. The ray tracer, once calibrated to the environment based on the limited measurements conducted, may be used to predict channel parameters at the reference points. Minor changes to the propagation environment may be quickly incorporated into the environment map utilized by the ray tracer, expediting the process of creating (and updating) the fingerprinting database. A neural network may be first trained on the synthetic data generated by the ray tracer, with the weights of the neural network refined by further training on real-world measurements.
\begin{figure}
	\centering
	\includegraphics[width=0.4\textwidth]{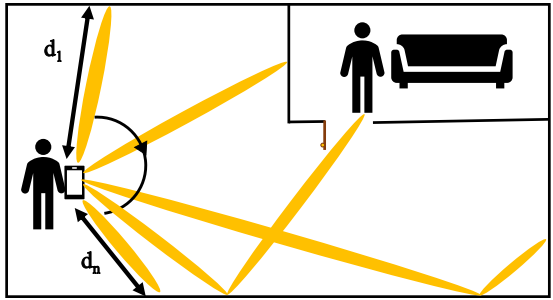}
	\caption{ Map generation on-the-fly and seeing through walls using narrow beam antennas and multipath. }
	\label{fig:mmwave_radar}
\end{figure}

\subsection{User Tracking and Data Fusion}\label{tracking}

Localization accuracy of a stationary target may be improved by averaging the position estimate, reducing the variance of the estimate. For mobile targets, the location must be estimated in a shorter period of time, which can be achieved via user tracking. User tracking refers to continuously estimating the position of a mobile UE, due to which sudden changes in the user's apparent position from one sampling instant to another, caused by positioning errors, may be smoothed out.

Modern cell phones are equipped with a variety of sensors. UEs possess an inertial measurement unit (IMU), consisting of a gyroscope to measure rotation, an accelerometer to measure acceleration, and a magnetometer to measure the magnetic field intensity. Given the initial position of the user, the current user position may be obtained by integrating the measured acceleration twice to get the user position. However, errors in IMU measurements grow with time - a constant offset in acceleration measurement leads to a quadratic error in position. 

Data from the sensors may be fused with channel measurement data using a Kalman filter/ extended Kalman filter (KF/EKF) to correct the drift in IMU measurements. A KF is a recursive linear estimator of the state (position and velocity) of a user. The current state of the user is modeled as a linear transformation of the state of the user at the previous time instant, based on kinematic equations derived from Newton's laws of motion, whereas sensor measurements are modeled as a linear transformation of the current state of the user. A KF is the optimal estimator of a linear process, given the mean and variance of the noise. If the relation is not linear, an EKF may be used to locally linearize the process via Taylor series expansion\cite{Pedersen_2018}. The KF/EKF minimizes the mean square error of the position estimate based on measurements obtained from all sensors up to the current time instant. When new information is obtained by the user in the form of new channel measurements or new sensor data, the KF/EKF recursively updates the position estimate based on the old position estimate and the new data.
\begin{table*}[]
	\centering
	\caption{Summary of the different position location techniques }\label{tbl:summary}
	\begin{tabular}{|c|c|c|c|c|}
		\hline
		\setlength\tabcolsep{1.5pt}
		\textbf{Position Location Method} & \textbf{Description} & \textbf{BS Density} & \textbf{Deployment Cost} & \textbf{Accuracy} \\ \hline
		\begin{tabular}[c]{@{}c@{}}Fundamental Techniques \\   \end{tabular}	 & \begin{tabular}[c]{@{}c@{}}Use uplink and downlink \\ AoA, ToA, TDoA measurements  \\to calculate position via geometry\end{tabular} & High & Low \cite{Kanhere18a}& Low\cite{Kanhere18a}  \\ 
		Cooperative Localization  	 & \begin{tabular}[c]{@{}c@{}}Use side-link (UE-UE) measurements to  \\ complement BS-UE measurements\end{tabular} & Low & Low \cite{Pedersen_2018} & Medium \cite{Pedersen_2018}\\ 
		Machine Learning & \begin{tabular}[c]{@{}c@{}}Channel features mapped to  \\ values stored in fingerprint database\end{tabular} & Medium & High \cite{Gante_2020} & High \cite{Gante_2020}\\ 
		User Tracking & \begin{tabular}[c]{@{}c@{}}Refine position estimate \\ of fundamental techniques,   \\ predict user trajectory with sensor data\end{tabular} & Medium & Low \cite{Pedersen_2018}& Medium \cite{Pedersen_2018}\\ 
		Multipath Exploiting Techniques & \begin{tabular}[c]{@{}c@{}}Extract position information\\  embedded in multipath components\end{tabular} & Low & Medium \cite{Rappaport19a,Shahmansoori2018}& High \cite{Rappaport19a,Shahmansoori2018}\\ \hline
	\end{tabular}
\end{table*}
\subsection{Localization Algorithms Exploiting Multipath }\label{MAP}
As discussed earlier, multipath components are conventionally thought to be a hindrance to accurate localization. However, in conjunction with a map of the environment, multipath components provide additional vital useful information regarding the location of the UE. For example, with a map of the environment available (Fig. 3), ``forbidden transitions" of a UE wherein the UE moves through walls or from one floor to another in consecutive time steps may be detected and discarded. 

 Multipath components from the BS may arrive at the UE via a direct path or via indirect paths along which the source ray suffers multiple reflections or scattering. Virtual anchors (VAs) are successive reflections of the BS on walls in the environment \cite{Pedersen_2018}, which are treated as an LoS BS in place of the physical NLoS BS. Future wireless devices will exploit real-time ray tracing \cite{Kanhere18a} for multipath propagation prediction in order to determine the VA locations. If the user's location is continuously tracked with an EKF, each multipath component received by the UE may be associated with a VA based on the previously estimated UE location. Once the correspondence between each multipath component and the VAs is known, any of the fundamental localization techniques (AoA, ToA, or TDoA) may be used to localize the UE.

With large bandwidths and narrow beamwidths at mmWave frequencies, more multipath components are resolvable, which makes the task of associating the multipath components with the VA more difficult. Ray tracing may be used to take advantage of NLoS multipath components arriving at a UE, providing single-shot user location estimation without user tracking. With knowledge of the AoA at the BS, the ToA of the source rays, and a map of the surrounding environment, the BS may determine the location of the UE via ray tracing each multipath component. Since it is not known whether the signal is reflected or transmitted through each obstruction along the traced signal path, two possible locations are recursively stored as ``candidate locations" at each obstruction encountered while ray tracing a multipath component. A majority of the candidate locations will be clustered near the true UE location, so the user may be localized to the centroid of the largest cluster of candidate locations \cite{Rappaport19a}.

In the absence of a map, with the assumption that each multipath component is reflected or scattered at most one time, the problem of determining the location of a UE can be reformulated into a nonlinear LS estimation problem \cite{Shahmansoori2018}. The scatterer/reflector positions and the UE position and orientation are estimated by jointly finding the scatterer and user locations where the expected distances and angles (geometrically calculated) match the measured distances and angles most closely in the least squared sense. Optimization techniques such as particle swarm optimization (PSO) and the LMA \cite{Shahmansoori2018} may be used for nonlinear LS optimization.

\section{Conclusion and Future Research}

This article has provided an overview of existing and emerging localization techniques, illustrating how utilizing the wide bandwidths at mmWave frequencies could lead to unprecedented localization accuracies. The narrow antenna beamwidths at mmWave frequencies require smart beam management, while optimal localization requires an exploration of multipath components arriving from all directions, for which a detailed study of joint communication and localization is required. Table \ref{tbl:summary} provides a summary of the different position location methods.

Looking into the future, we predict that a combination of machine learning, data fusion of measurements from multiple sensors, and cooperative localization will be used for robust, accurate position location. The wireless systems will need to seamlessly transfer the localization responsibility from one wireless technology (e.g., WiFi access points indoors) to another (e.g., cellular BSs outdoors), similar to handovers in current cellular networks when a user moves in and out of BS coverage cells. 

With centimeter-level localization accuracy in future cellular networks, privacy will become a growing concern. Users must be allowed to opt out of tracking if they so desire, and any user location data stored in the network must be protected from hackers. Additionally, the localization solution must be robust to interference from malicious users, who could, for instance, attempt to replicate the reference signals transmitted by the cellular network in order to gain unauthorized access to user location information.

The computing capabilities of UEs will enable mapping and ray tracing in real time. We envisage that cell phones in the future shall generate a map of the environment on the fly or have maps loaded within, enabling map-based localization algorithms that exploit real-time multipath propagation. The augmentation of human and computer vision will allow users to see in the dark and see through walls \cite{Rappaport19a}. Cell phones in the future will have the capability to either download or generate a map of the environment on the fly and ``see in the dark" \cite{Rappaport19a}. The UE will behave like a radar, measuring the distances of prominent features in the environment, such as walls, doors, and other obstructions. Additionally, reflections and scattering off walls will enable cell phones to view objects around corners or behind walls \cite{Rappaport19a}, as illustrated in Fig. \ref{fig:mmwave_radar}.

For ranging measurements, a radar operates in the \textit{pulsed radar} mode, wherein the radar transmits a single pulse, switches from transmit to receive mode, and waits for the echo back from the object that is to be range-estimated. However, due to constraints on switching speed, only objects at a sufficient distance from the user may be ranged. For example, an mmWave phased array with a TX-RX switching time of $ \sim $100 ns cannot range objects closer than 50 ft (electromagnetic waves travel 1 ft/ns). To range closer objects, a UE must simultaneously transmit and receive the radar signal, operating in the \textit{full duplex} mode, requiring TX-RX isolation \cite{Zhou_2017}. 

\section{Acknowledgments}
This work is supported by the NYU WIRELESS Industrial Affiliates Program and National Science Foundation (NSF) Grants 1702967, 1731290, and 1909206.

\section*{Biographies}

	\textsc{Ojas Kanhere} received the B.Tech. and M.Tech. degrees in electrical engineering from IIT Bombay, Mumbai, India, in 2017. He is currently pursuing the Ph.D. degree in electrical engineering with the NYU WIRELESS Research Center, New York University (NYU) Tandon School of Engineering, Brooklyn, NY, USA, under the supervision of Prof. Rappaport. His research interests include mmWave localization and channel modeling.

	\bigskip
	\noindent\textsc{Theodore S. Rappaport} (S'83-M'84-SM'91-F'98) is the David Lee/Ernst Weber Professor at New York University (NYU) and holds faculty appointments in the Electrical and Computer Engineering department, the Courant Computer Science department, and the NYU Langone School of Medicine. He founded NYU WIRELESS, a multidisciplinary research center, and the wireless research centers at UT Austin (WNCG) and Virginia Tech (MPRG).  His research has provided fundamental knowledge of wireless channels used to create the first Wi-Fi standard (IEEE 802.11), the first US digital TDMA and CDMA standards, the first public Wi-Fi hot spots, and more recently proved the viability of millimeter wave and sub-THz frequencies for 5G, 6G, and beyond.

\end{document}